\documentclass[prl,aps,twocolumn,superscriptaddress]{revtex4-2}
\usepackage{graphicx}
\usepackage{dcolumn}
\usepackage{bm}
\usepackage{lipsum}
\usepackage{ragged2e}
\usepackage{braket}
\usepackage{bbm}
\usepackage{lipsum}
\usepackage{color}
\usepackage{xcolor}
\usepackage{amsfonts}
\usepackage{amsmath}
\usepackage{amsmath,amssymb,latexsym}
\usepackage{wasysym,stmaryrd,marvosym}
\usepackage{mathrsfs}
\usepackage{soul}
\usepackage{dsfont}
\usepackage{float}

\usepackage[mathlines]{lineno}
\usepackage[colorinlistoftodos]{todonotes}
\usepackage[colorlinks=true, allcolors=blue]{hyperref}

\begin{document}

\newcommand{\ohio}{Department of Physics and Astronomy and Nanoscale and Quantum Phenomena Institute, Ohio University, Athens, Ohio 45701}

\title{Tilted Dirac cones and topological transitions in strained kagome lattices}

\author{M. A. Mojarro}
\email{mm232521@ohio.edu}
\author{Sergio E. Ulloa}
\affiliation{\ohio}

\date{\today}

\begin{abstract}
We study effects of strain on the electronic properties of the kagome lattice in a tight-binding formalism with spin-orbit coupling (SOC). The degeneracy
at the $\Gamma$ point evolves into a pair of emergent
tilted Dirac cones under uniaxial strain, where the anisotropy and tilting of the bands
depend on the magnitude and direction of the strain field. SOC opens gaps at the emergent Dirac points, making the flatband topological, characterized by a nontrivial $\mathbb{Z}_2$ index. Strains of a
few percent drive the system into trivial or topological phases. This
confirms that moderate strain can be used to engineer anisotropic Dirac bands with tunable 
properties to study new  phases in kagome lattices.


\end{abstract}


\maketitle

Two-dimensional (2D) kagome lattice symmetries have been studied since mid last century for Ising spins \cite{first}. Hopping electrons in this lattice are shown to result in graphene-like massless Dirac fermions and van-Hove singularities, as seen experimentally in the antiferromagnetic metal FeSn \cite{firstKagome}. The presence especially of a flatband in such lattice favors strong electron-electron interactions, and the competition between van-Hove singularities and Dirac and flatbands can lead to novel unexpected phenomena further assisted by the coexistence of topology and correlations \cite{review,van2}.

The recently discovered family of superconductors AV$_3$Sb$_5$ (A$=$K, Cs, Rb), contains vanadium atoms in a kagome plane \cite{van1,van2}. These compounds exhibit chiral charge order \cite{ChargeOrder,ChargeOrder2,ChargeOrder3} and unconventional superconducting pairing \cite{pairing}, suggesting a crucial connection between correlations and topology. For instance, it is found that pressure modulates the competition between superconductivity and charge order \cite{Pressure,TunableSup}, as well as shifting van-Hove singularities in this kind of materials \cite{VanHove}.
In addition to the AV$_3$Sb$_5$ family,
different strongly-correlated states are possible in a related Dirac-kagome herbertsmithite metal \cite{DiracMet}. 
{\em Ab initio} calculations also reveal the existence of a closely related family of kagome superconductors with promising rich behavior \cite{AbInitioKagome}.

The role of spin-orbit coupling (SOC) on kagome lattices has also been studied. A quantum spin Hall state with a $\mathbb{Z}_2$ topological index is achieved in the presence of SOC \cite{topo2}, and the finite Berry curvature leads to anomalous Hall effects \cite{topo1}. SOC combined with lattice dimerization allow tuning of the topological invariant at different fillings \cite{BreathKagome}, while metallic phases become possible when considering different on-site energies \cite{Z2Kagome}. 
Topological equivalence between Lieb and kagome lattices was shown under large strains \cite{LiebKagome} or large hopping modulations \cite{LiebK}. Similarly, strain was shown to modify the optical conductivity of 2D kagome lattices \cite{StrainTopo}.

The dispersionless band and Dirac points of kagome systems are protected by lattice symmetries suggesting their tunability by specific perturbations. Here we analyze the role of strain, SOC and site asymmetries on the topological properties of strained kagome lattices and show that it is possible to control the topological invariant near the 2/3 filling energy gaps by varying the magnitude and direction of uniaxial strain.  The creation of complex Berry phase structures in the system would further enhance the role of correlations in the superconducting or other interacting phases, and enrich the behavior seen in experiments.

\begin{figure*}
    \centering
    \includegraphics[scale=0.5]{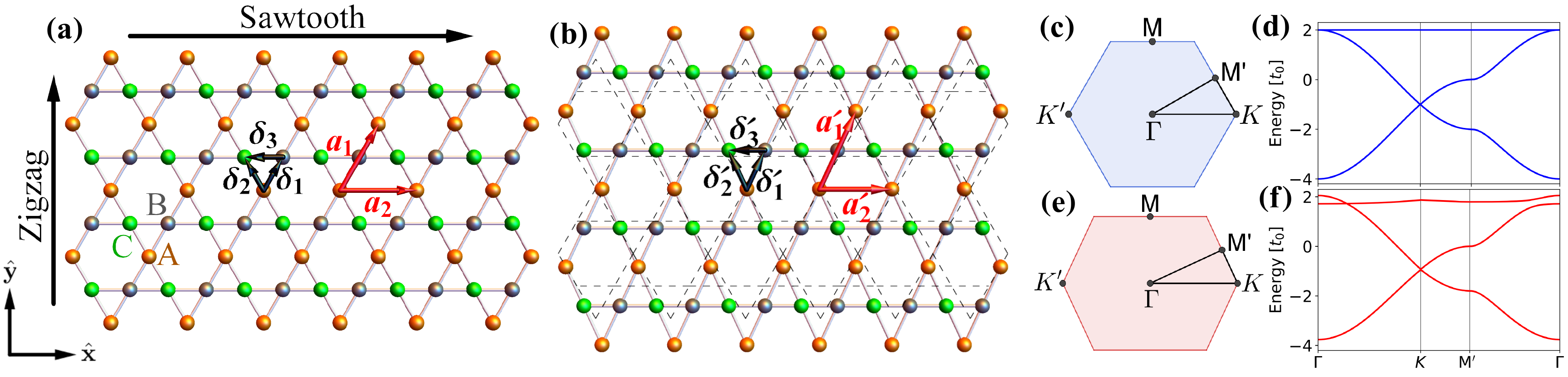}
    \caption{(a) Pristine kagome, and (b) strained kagome lattice along the zigzag direction. Nearest-neighbor vectors (black), as well as primitive vectors (red) are shown. (c) and (d) show the band structure of the unstrained kagome lattice and corresponding Brillouin zone path. (e) and (f) show the spectrum of the zigzag-strained lattice and corresponding Brillouin zone path.}
    \label{fig1}
\end{figure*}

We use a tight-binding description of the electronic properties of kagome lattices as it compares well with DFT calculations and experimental characterization \cite{firstKagome}. The kagome tripartite crystal structure is defined by three basis vectors connecting nearest-neighbor sites: $\bm{\delta}_1=a_0(1/2,\,\sqrt{3}/2)$, $\bm{\delta}_2=a_0(-1/2,\,\sqrt{3}/2)$, and $\bm{\delta}_3=\bm{\delta}_2-\bm{\delta}_1$ (with $a_0$ the inter-atomic distance), and two primitive vectors of the triangular Bravais lattice: $\textbf{a}_1=2\,\bm{\delta}_1$ and $\textbf{a}_2=-2\,\bm{\delta}_3$, as shown in Fig.\ \ref{fig1}(a). Strain deforms the two-dimensional lattice, as characterized by a displacement field $\mathbf{u}(\mathbf{r})=(u_x(\textbf{r}),\,u_y(\textbf{r}))$, where $\mathbf{r}$ is the atomic position vector. The corresponding deformed lattice sites are at $\mathbf{r}'=\mathbf{r}+\mathbf{u}(\mathbf{r})$. For uniform strain, the displacement field is $\mathbf{u}(\mathbf{r})=\hat{\epsilon}\cdot\mathbf{r}$, with the strain tensor $\hat{\epsilon}$ given in terms of the Poisson ratio $\rho$ as \cite{tensor}
\begin{equation}
    \hat{\epsilon}=
    \begin{pmatrix}
    \epsilon(\cos^2\theta-\rho\sin^2\theta) & \epsilon(1+\rho)\cos\theta\,\sin\theta\\
    \epsilon(1+\rho)\cos\theta\,\sin\theta & \epsilon(\sin^2\theta-\rho\cos^2\theta)
    \end{pmatrix}\,,
\end{equation}
where $\epsilon$ denotes the strain magnitude and $\theta$ its direction with respect to the $x$ axis ($\textbf{a}_2$ direction). The nearest-neighbor vectors transform as $\bm{\delta}'_i=(\mathds{1}+\hat{\epsilon})\cdot\bm{\delta}_i$, where $\mathds{1}$ is a $2\times2$ identity matrix. This changes the hopping integral between nearest-neighbor sites $i$ and $j$:
\begin{equation}
    t_{ij}=t_0\,\text{exp}\left[-\beta\left(\frac{|\bm{\delta}_{ij}'|}{a_0}-1\right)\right],
\end{equation}
where $t_0$ is the hopping in the absence of strain, $\beta$ the Grüneisen parameter, and $\bm{\delta}_{ij}$ the vector between sites $i$ and $j$ \footnote{For numerics, we choose parameters as in graphene, $\beta\approx3$ and $\nu\approx0.165$ \cite{grun}, as corresponding values for kagome materials have not yet been reported.}.

The Hamiltonian of the strained kagome lattice is then
\begin{eqnarray}
    \mathcal{H}_{0}&=&-\sum_{\langle ij\rangle\sigma} t_{ij} h^\dagger_{i\sigma}h_{j\sigma}+\sum_{i,\sigma}\varepsilon_{i}h_{i\sigma}^\dagger h_{i\sigma}\,,
\end{eqnarray}
where $h_{i\sigma}^\dagger$ ($h_{i\sigma}$) creates (annihilates) a particle at site $i=$ A, B, C with spin $\sigma$ (up or down) and $\langle ij\rangle$ denotes sum over first-neighbor sites. Different on-site sublattice energies $\varepsilon_i$ represent different atomic species or chemical environment.


The SOC Hamiltonian takes the form \cite{topo2,kane1}
\begin{eqnarray}
\nonumber \mathcal{H}_{\text{SOC}}&=&i\lambda_{I}\sum_{\langle\langle ij\rangle\rangle \sigma\sigma'}\nu_{ij}h_{i\sigma}^\dagger s^z_{\sigma\sigma'}h_{j\sigma'}\\
&&+\,i\lambda_{R}\sum_{\langle ij\rangle \sigma\sigma'}h_{i\sigma}^\dagger\left(\textbf{s}_{\sigma\sigma'}\times\bm{\delta}'_{ij}\right)_z h_{j\sigma'}\,,
\end{eqnarray}
where $\langle\langle ij \rangle\rangle$ denotes sum over second-neighbor sites, $\textbf{s}=(s^x,s^y,s^z)$ is the vector of Pauli matrices acting on spin space, and $\nu_{ij}=\text{sgn}(\bm{\delta}'_{ik}\times\bm{\delta}'_{kj})_z$. The $\lambda_I$ term accounts for the intrinsic SOC between second-neighbor sites which respects all symmetries and drives the system into a quantum spin Hall state \cite{kane1}. First-neighbor SOC is also allowed by symmetry in this system \cite{BreathKagome}, and can be further considered \cite{Supl}. 
The Rashba SOC (strength $\lambda_R$), is associated with broken inversion symmetry, as that provided by an out-of-plane electric field \cite{RashbaSHE,RashbaSHE2}. The total Hamiltonian in momentum space can be written as $\mathcal{H}=\sum_{\bf{k}}d^{\dagger}_{\bf{k}}H({\bf k})d_{\bf{k}}$, where $d_{\bf k}=(a_{{\bf k},\uparrow},\,b_{{\bf k},\uparrow},\,c_{{\bf k},\uparrow},\,a_{{\bf k},\downarrow},\,b_{{\bf k},\downarrow},\,c_{{\bf k},\downarrow})^{\text{T}}$ and $H({\bf k})$ is given by 
\begin{eqnarray}\label{BlockH}
    \nonumber H(\textbf{k})&=&-2\sum_{i=1}^3t_i\cos\left(\textbf{k}\cdot\bm{\delta}'_i\right)\,s^0\otimes S_i+\sum_{i=4}^6\varepsilon_{i}\,s^0\otimes S_i\\
    \nonumber&&+\,i2\lambda_I\sum_{i=1}^3 \cos(\textbf{k}\cdot\tilde{\bm{\delta}}'_i)\,s^z\otimes A_i\\
    &&-\,2\lambda_R\sum_{i=1}^3\sin\left(\textbf{k}\cdot\bm{\delta}'_i\right)\,\left(\textbf{s}\times\bm{\delta}_i'\right)_z\otimes S_i\,,
\end{eqnarray}
where $\textbf{k}=(k_x,\,k_y)$ is the electron wave vector, $\varepsilon_{4,5,6}=\varepsilon_{\text{A}, \text{B}, \text{C}}$, $s^0$ is the identity matrix in spin space, and we have defined $\tilde{\bm{\delta}}'_1=\bm{\delta}'_2+\bm{\delta}'_3$, $\tilde{\bm{\delta}}'_2=\bm{\delta}'_1-\bm{\delta}'_3$ and $\tilde{\bm{\delta}}'_3=\bm{\delta}'_1+\bm{\delta}'_2$. The sets $\{S_i\}$ and $\{A_i\}$ form a basis of $3\times3$ symmetric and antisymmetric (skew-symmetric) matrices, respectively \cite{Supl}.

\begin{figure}
    \centering
    \includegraphics[scale=0.245]{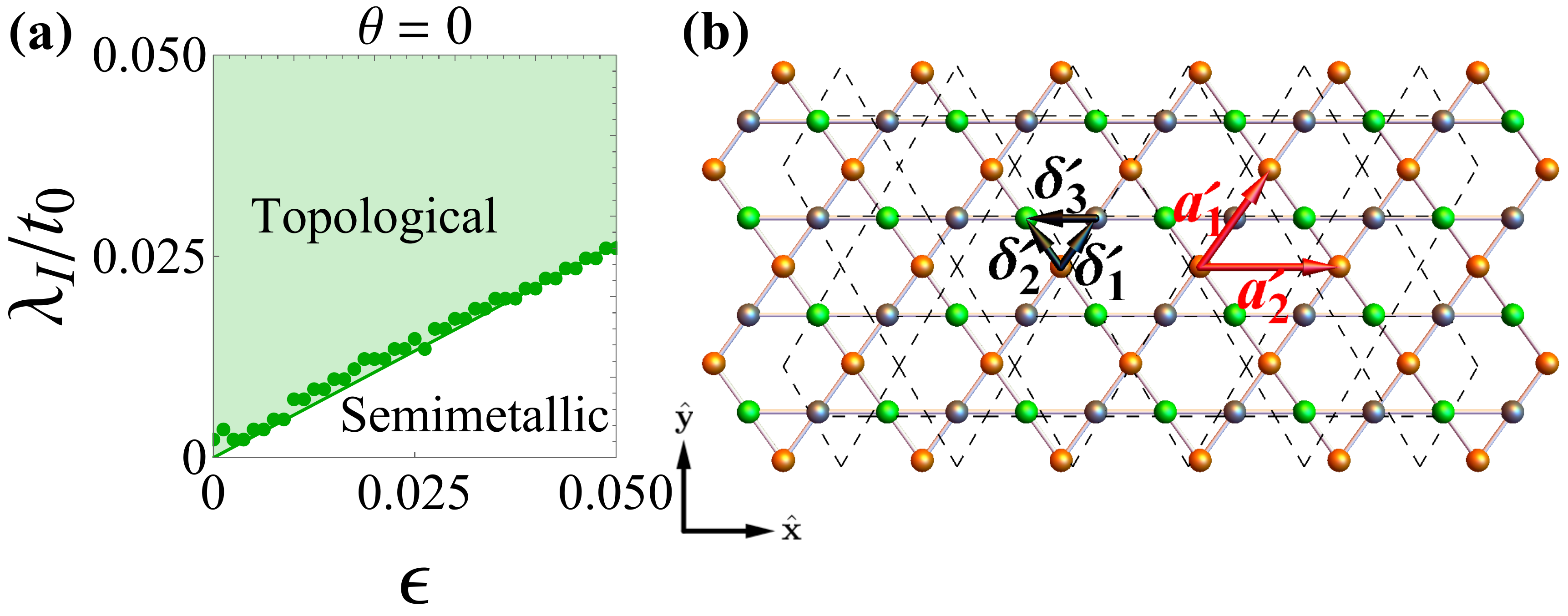}
    \caption{(a) Topological phase diagram at $2/3$ filling as function of the intrinsic SOC and strain along the sawtooth direction ($\theta=0$, as shown in (b)) with $\lambda_R=0.2t_0$, and all same on-site energies. Green region indicates topologically nontrivial phase and white the semimetallic phase. For fixed intrinsic SOC, an increasing deformation makes the system semimetallic, closing the gap between the upper bands. Green dots correspond to numerical calculations of bandgap vanishing. The straight line is a simple fitting.}
    \label{fig2}
\end{figure}

In the pristine system, with the same on-site energies and in the absence of SOC or strain, the well-known spectrum is shown in Fig.\ \ref{fig1}(d). At the $\Gamma$ point, the parabolic and upper flatband are degenerate, while  massless Dirac fermions describe the vicinity of the $K$, $K'$ points located at $(\pm 2\pi/3a_0,\,0)$, respectively. 
Applying strain shifts the Dirac points by a vector potential similar to graphene \cite{strainKagome,Supl}, while the degeneracy at the $\Gamma$ point splits into two Dirac points $Q_{\eta}$ ($\eta=\pm$) with location in momentum space that depends on the magnitude and direction of strain \cite{Supl}. The dispersion in the vicinity of the emerging Dirac points is given by tilted Dirac cones with a two-level Hamiltonian of the form \cite{Supl} $H_\theta^\eta(\textbf{q})=\varepsilon^\eta_\theta(\textbf{q})\mathds{1}+\bm{\sigma}\cdot\textbf{d}^\eta_\theta(\textbf{q})$, where $\textbf{q}=(q_x,\,q_y)$ is the momentum measured relative to $Q_{\eta}$, and $\bm{\sigma}$ is a vector of Pauli matrices.
For strain along the sawtooth direction ($\theta=0$, see Fig.\ \ref{fig2}(b)), we find $\textbf{d}_0^\eta(\textbf{q})=\eta(-v_x\hbar q_x,\,0,\,v_y\hbar q_y)$ and $\varepsilon_0^\eta(\textbf{q})=2t_0(1-\beta\epsilon)-\eta v_t\hbar q_y$. The velocities $v_i$ are determined by the strain parameters $\epsilon$, $\nu$, $\beta$ \cite{Supl}. In this case, the cone tilting is along the $q_y$ axis and characterized by  $\gamma=v_t/v_y$. 
For strains along any sawtooth direction of the lattice, we find that $\gamma=1$, which describes a type-III Dirac band with critical tilting \cite{weyl,type3}. If the strain is applied along the zigzag direction ($\theta=\pi/2$, see Fig.\ \ref{fig1}(b)), we find $\textbf{d}_{\pi/2}^\eta(\textbf{q})=-\eta(v_y\hbar q_y,\,0,\,v_x\hbar q_x)$ and $\varepsilon_{\pi/2}^\eta(\textbf{q})=2t_0(1-\beta\epsilon)-\eta v_t\hbar q_x$. Here, the tilting is along the $q_x$ axis and characterized by $v_t/v_x<1$, describing tilted type-I Dirac cones \cite{weyl,type3}.  

Interestingly, inversion symmetry is not broken for different on-site energies, and the degeneracy between the lower bands is preserved in the kagome lattice, as the Dirac points are only displaced for varying site asymmetries \cite{topo2}. 
For $\varepsilon_{\text{A}}>0$ and $\varepsilon_{\text{B}}=\varepsilon_{\text{C}}=0$, it can be shown from the eigenvalues of Eq.\ \eqref{BlockH} (in the absence of strain and SOC) that the Dirac points
are relocated to $(\pm \kappa_x,\,0)$, where $4\cos(a_0 \kappa_x)=1+\varepsilon_{\text{A}}/t_0-\sqrt{8+(\varepsilon_{\text{A}}/t_0+1)^2}$ (or $\kappa_x\approx2\pi/3a_0-\varepsilon_{\text{A}}/\left(3\sqrt{3}t_0a_0\right)$ for $\varepsilon_{\text{A}}/t_0\ll1$).

Similarly, site asymmetries shift the degeneracy at the $\Gamma$ point,
which is relocated to $(0,\pm \kappa_y)$, where $\cos(\sqrt{3}a_0\kappa_y)=1-\varepsilon_{\text{A}}/t_0$ (or $\kappa_y\approx\sqrt{2\varepsilon_{\text{A}}/(3t_0a_0^2)}$ for $\varepsilon_{\text{A}}/t_0\ll1$).

The inclusion of intrinsic SOC gaps the two lower bands at the Dirac points and topological conducting states manifest at the edges \cite{kane1,kanez2}. The degeneracy between the higher bands is also lifted and the gap opening results in the system having a non-trivial topology. The Rashba interaction breaks the single cone degeneracy at $K$, $K'$, following the triangular symmetry of the lattice as in graphene \cite{nancyRSO}. 

To fully characterize the topological properties of the strained kagome lattice we calculate the $\mathbb{Z}_2$ topological invariant $\nu$, tracking hybrid Wannier charge centers using the Z2pack \cite{z2pack}, and studying the edge states of strips of kagome lattice. The $\mathbb{Z}_2$ index agrees with parity eigenvalue results for inversion symmetric cases \cite{FuMele}. Whenever $\nu=1$, the system supports topological edge states, while remaining insulating in the bulk. If $\nu=0$, the system is in a conventional (trivial) insulating state. 

At $1/3$ filling, topological edge states become possible as long as the Rashba coupling is weak enough, 
$3\lambda_R<\sqrt{3}t_0 + 6 \lambda_I - \sqrt{3t_0^2 + 12 \lambda_I^2}$. In the opposite regime, 
the system becomes semimetallic and the energy gap at $K$, $K'$ vanishes. In this sense, the topological character at 1/3 filling exhibits graphene-like behavior \cite{kanez2}.

We now focus on the system at 2/3 filling. In the absence of strain and for the same on-site energies, the gap at the $\Gamma$ point has a magnitude of $4\sqrt{3}\lambda_I$, independent of the Rashba interaction, as one can see from Eq.\ \eqref{BlockH}. Correspondingly, the topological nature of the bands in the presence of intrinsic SOC appears robust against such inversion-symmetry breaking perturbation. 

\begin{figure*}
    \centering
    \includegraphics[scale=0.29]{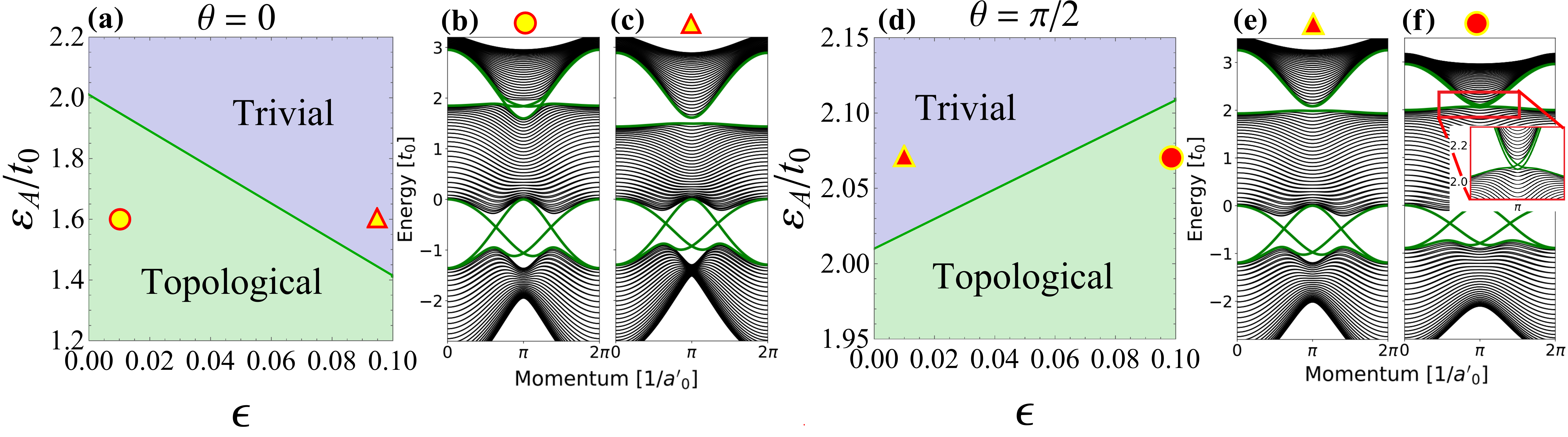}
    \caption{Topological phase diagram at 2/3 filling (energy $\simeq 2t_0$) for $\lambda_I=0.1\,t_0$ and $\lambda_R=0$ as function of the on-site energy $\varepsilon_\text{A}$ ($\varepsilon_\text{B}=\varepsilon_\text{C}=0$) and strain magnitude $\epsilon$ along the (a) sawtooth ($\theta=0$) and (d) zigzag ($\theta=\pi/2$) directions. 
    Increasing strain along the sawtooth direction produces a transition from a topological (red region) to a trivial (blue region) insulating phase near energy $2t_0$. This is validated by edge states in an infinite strip of kagome lattice along the $\textbf{a}_1$ direction with 30 unit cells along $\textbf{a}_2$. For parameters in the topological region in (a), conducting edge states in green in (b) appear midgap (yellow circle). In the trivial region, there are no midgap edge states, as shown in (c) (yellow triangle). Increasing strain in the zigzag direction drives the system from a trivial to a topological phase as shown in (d), revealed as well by the appearance of midgap edge states in (e) (red triangle) and (f) (red circle). Notice edge states near 1/3 filling in all cases shown.}
    \label{fig3}
\end{figure*}

\begin{figure}
    \centering
    \includegraphics[scale=0.39]{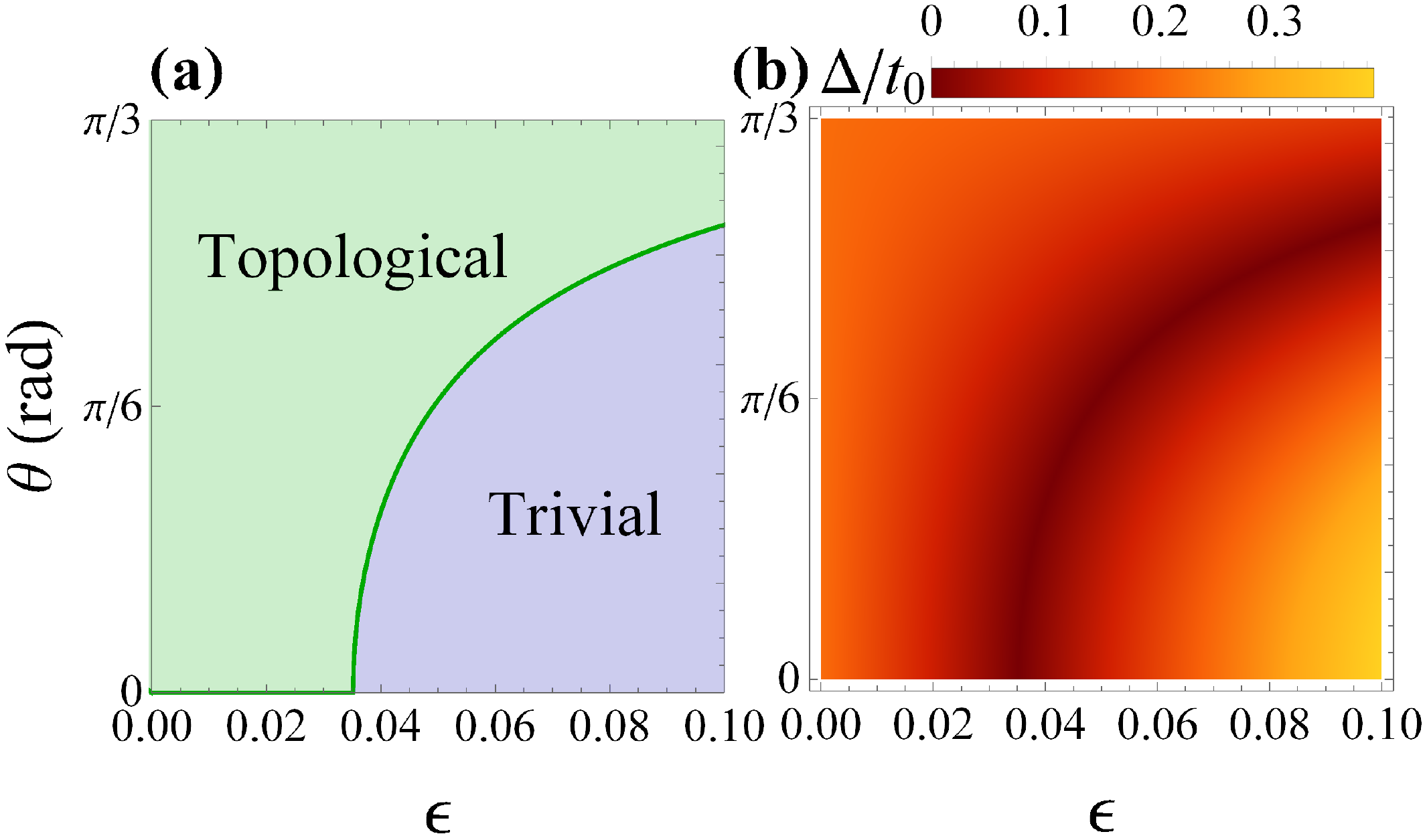}
    \caption{Topological phase diagram at $2/3$ filling with $\lambda_I=0.1t_0$, $\lambda_R=0$, and $\varepsilon_{\text{A}}=1.8t_0$, as function of the magnitude $\epsilon$ and direction $\theta$ of the strain. (b) shows magnitude of the gap $\Delta$ at the M point over the same region \cite{Supl}.}
    \label{fig4}
\end{figure}

Most importantly, a uniform strain allows one to tune the topological invariant at $2/3$ filling. When sufficiently strong, the strain drives the system into either semimetallic, trivial ($\nu=0$), or topological ($\nu=1$) phases, depending on structure parameters of the system.
When the on-site energies are the same, the strain drives the system into a semimetallic phase for fixed Rashba and intrinsic SOC, as shown in Fig.\ \ref{fig2}, producing a bandgap closing between the upper and middle bands.

For the site-asymmetric case $\varepsilon_{\text{A}}>\varepsilon_{\text{B}}=\varepsilon_{\text{C}}=0$, with fixed $\lambda_I$ and $\varepsilon_{\text{A}}$, and $\lambda_R=0$, increasing strain produces bandgap closing that occurs at the M point of the strained Brillouin zone (see Fig.\ \ref{fig1}(e)). 
For strain along the sawtooth direction, the M point is relocated to $\left(0,\,\pi/\left[\sqrt{3}a_0(1-\rho \epsilon)\right]\right)$, and the energy gap between the upper and middle bands reads $\left|\varepsilon_{\text{A}}-2\sqrt{\lambda_I^2+t_0^2(\beta\epsilon-1)^2}\right|$ \cite{Supl}. As strain closes the gap, a topological transition to a trivial insulator phase occurs--Fig.\ \ref{fig3}(a). A strip of kagome lattice extended along $\textbf{a}_1$ in the topological regime exhibits well-defined midgap edge states, while the bulk remains gapped, as shown in Fig.\ \ref{fig3}(b). Such edge states disappear when in the trivial phase, Fig.\ \ref{fig3}(c).
For strain along the zigzag direction, the energy gap is given by $\left|\varepsilon_{\text{A}}-2\sqrt{\lambda_I^2+t_0^2(\beta\rho\epsilon+1)^2}\right|$ at the M point, located now at $\left(0,\,\pi/\left[\sqrt{3}a_0(1+ \epsilon)\right]\right)$ \cite{Supl}.
In contrast to the sawtooth-strain case, the system can be driven here from a trivial to a topological phase at fixed $\lambda_I$ and $\varepsilon_{\text{A}}$ by increasing strain, as shown in Fig.\ \ref{fig3}(d). The associated edge states reveal the trivial (Fig.\ \ref{fig3}(e)) and topological (\ref{fig3}(f)) nature of the system at 2/3 filling.

Figure \ref{fig4}(a) shows the phase diagram as a function of magnitude and direction of the strain while keeping $\varepsilon_\text{A}$ and $\lambda_I$ constant. As before, the topological phase transition is accompanied by a bandgap closing at the M point of the strained Brillouin zone \cite{Supl}. The magnitude of the gap between the upper bands is shown in Fig.\ \ref{fig4}(b) for the same parameters. As expected, the band gap vanishes at the border between topological and trivial phases.

\begin{figure}
    \centering
    \includegraphics[scale=0.37]{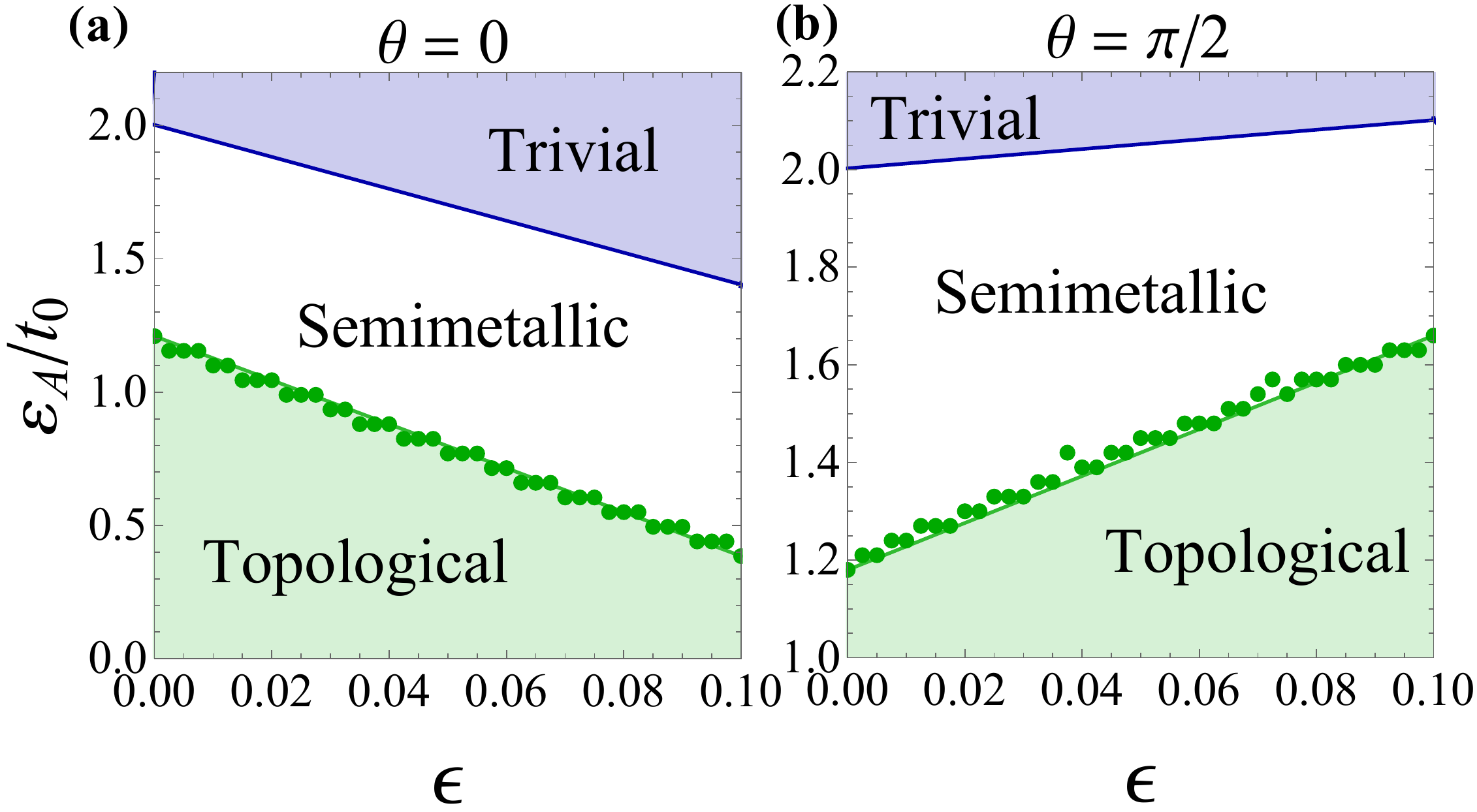}
    \caption{Phase diagram at $2/3$ filling as function of strain magnitude and on-site energy $\varepsilon_{\text{A}}$ with $\lambda_I=0.05t_0$, $\lambda_R=0.1t_0$. The strain is applied along the (a) sawtooth and (b) zigzag directions.}
    \label{fig5}
\end{figure}

Once the Rashba interaction is included, a semimetallic phase appears between trivial and nontrivial gapped phases. For strain along the sawtooth direction, a sufficiently small strain and/or $\varepsilon_\text{A}$ keeps the system in a topological phase, Fig.\ \ref{fig5}(a). Increasing either makes the system semimetallic, as the energy gap between the upper bands vanishes. Further increases result in the system eventually reaching a trivial insulating phase, as the band gap at $2/3$ filling reopens. Similar phase changes can also be produced when the strain is applied along the zigzag direction, as shown in  Fig.\ \ref{fig5}(b).

We should mention that the topological properties at $1/3$ filling depend only on the strength of the SOC, and are robust against external deformations. In contrast to graphene, we also see that on-site energy differences do not induce topological phase transitions at $1/3$ filling in kagome lattices.

We have shown that moderate strains can modify the topological properties of electronic states in kagome lattices. Through analytical and numerical calculations of the $\mathbb{Z}_2$ topological index and analysis of edge states, we can specify the conditions for topological behavior. The presence of intrinsic SOC makes the system topological at 2/3 filling with nearly dispersionless bulk bands and well-defined edge states. As the Rashba interaction is included, the system can be driven into a semimetallic phase when the strain is sufficiently strong. Different topological and trivial phases are also tunable depending on the direction and magnitude of the strain. Such control with moderate strain may allow the study of interesting tunable transport behavior, as edge and bulk conduction compete with one another and result in different magnetotransport response \cite{Culcer_2020}. Similarly, the tunable onset of drastically different topology near van-Hove singularities may result in interacting many body states with corresponding strain-controlled properties \cite{FQHE_2011}.

We thank discussions with N. Sandler and E. Vernek. Supported by U.S. Department of Energy, Office of Basic Energy Sciences, Materials Science and Engineering Division.

\bibliography{biblio.bib}
\end{document}


\newcommand{\cnyn}{Centro de Nanociencias y Nanotecnolog\'ia,
Universidad Nacional Aut\'onoma de M\'exico, Apartado Postal 2681, 22800, Ensenada, Baja California, M\'exico.}
\newcommand{\uabc}{Facultad de Ciencias, Universidad Aut\'onoma de Baja California, Apartado Postal 1880, 22800 Ensenada, Baja California, M\'exico}
\newcommand{\ohio}{Department of Physics and Astronomy and Nanoscale and Quantum Phenomena Institute, Ohio University, Athens, Ohio 45701}

\title{Tilted Dirac cones and topological transitions in strained kagome lattices}

\author{M. A. Mojarro}
\author{Sergio E. Ulloa}
\affiliation{\ohio}


\maketitle

\begin{center}
    {\bf{SUPPLEMENTAL MATERIAL}}
\end{center}

\vspace{.3cm}

The effect of a uniform strain on the energy spectrum of the kagome lattice is shifting the degeneracy at the $\Gamma$ point into two degenerate points. Here, we show that the Hamiltonian in the vicinity of such points describes massless Dirac fermions. Analytical expressions of the energy gap at the M point of the strained Brillouin zone are given, where the band gap closing-opening mechanism accompanies topological phase transitions. We also discuss how the energy gap at the Dirac points is independent of the on-site energies, and it only depends on the strength of the spin-orbit coupling (SOC). Finally, we briefly discuss the effect of a first-neighbors intrinsic SOC on the energy spectrum.

\section{Momentum space Hamiltonian}

The main text shows the Bloch   Hamiltonian in momentum space of a strained kagome lattice with SOC after performing a Fourier transform to Eqs.(3)-(4)
\begin{equation}\label{s1} H(\textbf{k})=-2\sum_{i=1}^3t_i\cos(\textbf{k}\cdot\bm{\delta}'_i)\,s^0\otimes S_i+\sum_{i=4}^6\varepsilon_{i}\,s^0\otimes S_i+\,i2\lambda_I\sum_{i=1}^3 \cos(\textbf{k}\cdot\tilde{\bm{\delta}}'_i)\,s^z\otimes A_i-\,2\lambda_R\sum_{i=1}^3\sin(\textbf{k}\cdot\bm{\delta}'_i)\,(\textbf{s}\times\bm{\delta}_i')_z\otimes S_i,
\end{equation}
where $\lambda_I$ and $\lambda_R$ are the intrinsic and Rashba SOC strengths, respectively, $\textbf{s}=(s^x,\,s^y,\,s^z)$ is the vector of Pauli matrices acting on spin space with $s^0$ the identity matrix, $\varepsilon_{4,5,6}=\varepsilon_{\text{A}, \text{B}, \text{C}}$ are on-site energies of the three sublattices of the kagome lattice, and we have defined $\tilde{\bm{\delta}}'_1=\bm{\delta}'_2+\bm{\delta}'_3$, $\tilde{\bm{\delta}}'_2=\bm{\delta}'_1-\bm{\delta}'_3$ and $\tilde{\bm{\delta}}'_3=\bm{\delta}'_1+\bm{\delta}'_2$. We have introduced a set of symmetric matrices given by   
\begin{equation}
    S_1=\begin{pmatrix}
        0 & 1 &0\\
        1 & 0 & 0\\
        0 & 0 & 0
    \end{pmatrix},\quad
     S_2=\begin{pmatrix}
        0 & 0 & 1\\
        0 & 0 & 0\\
        1 & 0 & 0
    \end{pmatrix},\quad
     S_3=\begin{pmatrix}
        0 & 0 &0\\
        0 & 0 & 1\\
        0 & 1 & 0
    \end{pmatrix},\quad
     S_4=\begin{pmatrix}
        1 & 0 & 0\\
        0 & 0 & 0\\
        0 & 0 & 0
    \end{pmatrix},\quad
     S_5=\begin{pmatrix}
        0 & 0 & 0\\
        0 & 1 & 0\\
        0 & 0 & 0
    \end{pmatrix},\quad
     S_6=\begin{pmatrix}
        0 & 0 & 0\\
        0 & 0 & 0\\
        0 & 0 & 1
    \end{pmatrix},
\end{equation}
as well as antisymmetric matrices 
\begin{equation}
    A_1=\begin{pmatrix}
        0 & 1 &0\\
        -1 & 0 & 0\\
        0 & 0 & 0
    \end{pmatrix},\quad
     A_2=\begin{pmatrix}
        0 & 0 & -1\\
        0 & 0 & 0\\
        1 & 0 & 0
    \end{pmatrix},\quad
     A_3=\begin{pmatrix}
        0 & 0 &0\\
        0 & 0 & 1\\
        0 & -1 & 0
    \end{pmatrix}.
\end{equation}
The hopping amplitudes $t_i$ and the transformed vectors $\bm{\delta}_i'$ 
connecting nearest-neighbor sites are defined in the manuscript. 

\section{Hamiltonian in the vicinity of the $\Gamma$ point and Dirac Hamiltonians}

\begin{figure}
    \centering
    \includegraphics[scale=0.33]{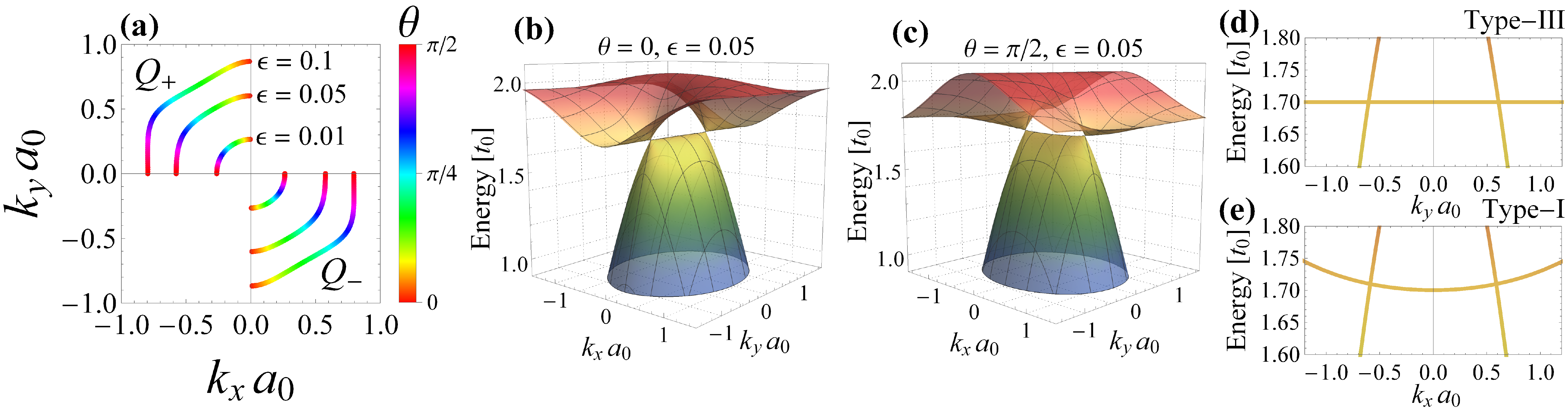}
    \caption{(a) Position of the degenerate points $Q_{\pm}$ in the first Brillouin zone as a function of the direction of strain for several values of the magnitude of strain $\epsilon$. In (b) and (c) we show the upper bands in the vicinity of the $\Gamma$ point for strain along the sawtooth ($\theta=0$) and zigzag ($\theta=\pi/2$) directions, respectively, and strain magnitude $\epsilon=0.05$. (d) and (e) show a cut of (b) and (c) along the $k_x$ and $k_y$ directions, respectively, showing the distinction between tilted type-I and type-III Dirac cones.}
    \label{fig1S}
\end{figure}

To study the vicinity of the $\Gamma$ point, we expand the Hamiltonian in Eq.\ \eqref{s1} in the basis formed by the eigenstates of the flat and middle bands at the $\Gamma$ point. In the absence of SOC and on-site energies, and before the strain is applied, we obtain

\begin{equation}
    H_{\Gamma}(\textbf{k})=2t_0\mathds{1}-
    t_0a_0^2\begin{pmatrix}
    k_x^2 & k_xk_y\\
    k_xk_y & k_y^2\\
    \end{pmatrix}=2t_0\mathds{1}-t_0a_0^2\begin{pmatrix}
    \textbf{k}\cdot\bar{\kappa}_1\cdot\textbf{k} & \textbf{k}\cdot\bar{\kappa}_3\cdot\textbf{k}\\
    \textbf{k}\cdot\bar{\kappa}_3\cdot\textbf{k} & \textbf{k}\cdot\bar{\kappa}_2\cdot\textbf{k}\\
    \end{pmatrix}
\end{equation}
where we have defined
\begin{equation}
    \bar{\kappa}_1=\begin{pmatrix}
    1 & 0\\
    0 & 0
    \end{pmatrix},\qquad \bar{\kappa}_2=\begin{pmatrix}
    0 & 0\\
    0 & 1
    \end{pmatrix},\qquad \bar{\kappa}_3=\frac{1}{2}\begin{pmatrix}
    0 & 1\\
    1 & 0
    \end{pmatrix}.
\end{equation}
The energy spectrum of the last Hamiltonian consists of a flat band and a parabolic band, as expected. Once a uniform strain is applied (space-independent), the last Hamiltonian takes the following form for small deformations ($\epsilon\ll1$)
\begin{equation}
    H_{\Gamma}(\textbf{k})=2t_0(\mathds{1}-\beta\hat{\epsilon})-t_0a_0^2\begin{pmatrix}
    \textbf{k}\cdot\bar{\kappa}'_1\cdot\textbf{k} & \textbf{k}\cdot\bar{\kappa}'_3\cdot\textbf{k}\\
    \textbf{k}\cdot\bar{\kappa}'_3\cdot\textbf{k} & \textbf{k}\cdot\bar{\kappa}'_2\cdot\textbf{k}\\
    \end{pmatrix},
\end{equation}
where $\beta$ is the Grüneisen parameter, $\hat{\epsilon}$ is the strain tensor:
\begin{equation}
  \hat{\epsilon}=\begin{pmatrix}
      \epsilon_{xx} & \epsilon_{xy}\\
      \epsilon_{xy} & \epsilon_{yy}
  \end{pmatrix}=
    \begin{pmatrix}
    \epsilon(\cos^2\theta-\rho\,\sin^2\theta) & \epsilon(1+\rho)\cos\theta\,\sin\theta\\
    \epsilon(1+\rho)\cos\theta\,\sin\theta & \epsilon(\sin^2\theta-\rho\,\cos^2\theta)
    \end{pmatrix},
\end{equation}
with $\epsilon$ the strain magnitude, $\theta$ the direction of the applied strain with respect to the horizontal, $\rho$ the Poisson ratio, and now
\footnotesize{
\begin{equation}
    \bar{\kappa}_1'= \bar{\kappa}_1-\begin{pmatrix}
    \epsilon_{xx}(\beta-2) & -\epsilon_{xy}\\
    -\epsilon_{xy} & 0
    \end{pmatrix},\qquad \bar{\kappa}_2'= \bar{\kappa}_2-\frac{1}{4}\begin{pmatrix}
    \beta(\epsilon_{yy}-\epsilon_{xx}) & 2\epsilon_{xy}(\beta-2)\\
    2\epsilon_{xy}(\beta-2) &\beta(\epsilon_{xx}+3\epsilon_{yy})-8\epsilon_{yy}
    \end{pmatrix},\qquad     \bar{\kappa}_3'=\bar{\kappa}_3-\frac{1}{4}\begin{pmatrix}
    \epsilon_{xy}(\beta-4) & \epsilon_{xx}(\beta-4)\\
    \epsilon_{yy}(3\beta-4) & \epsilon_{xy}(3\beta-4)
    \end{pmatrix}.
\end{equation}
}
\normalsize
The degeneracy at the $\Gamma$ point splits into two degenerate points, $Q_{\eta}$ ($\eta=\pm$), once the strain is applied. The location of such degenerate points is given by $\textbf{Q}_{\eta}(\epsilon,\,\theta)
=\eta\, Q(\phi)\,(\cos\phi,\,\sin\phi)$, where
\begin{eqnarray}
Q(\phi)&=&\frac{1}{a_0}\sqrt{\frac{8\beta\epsilon(1+\rho)}{\epsilon(1+\rho)[2\beta-4+\beta\cos(4\theta+2\phi)]-2[2+\epsilon(\beta-2)(\rho-1)]\cos(2\theta-2\phi)}}\label{Q}\,,\\
\tan2\phi&=&\frac{4+2\epsilon(\beta-2)(\rho-1)\sin2\theta-\beta\epsilon(1+\rho)\sin4\theta}{4+2\epsilon(\beta-2)(\rho-1)\cos2\theta+\beta\epsilon(1+\rho)\cos4\theta}\,.
\end{eqnarray}
Figure \ref{fig1S} shows the position of the degenerate points $Q_{\pm}$ in momentum space for several values of the strain magnitude $\epsilon$ as a function of the strain direction $\theta$.

The Hamiltonian in the vicinity of the new degenerate points is given by a two-level form $H_\theta^\eta(\textbf{q})=\varepsilon^\eta_\theta(\textbf{q})\mathds{1}+\bm{\sigma}\cdot\textbf{d}^\eta_\theta(\textbf{q})$, where $\textbf{q}=(q_x,\,q_y)$ is the momentum measured relative to the degenerate points. We have introduced a set of Pauli matrices $\bm{\sigma}=(\sigma^x,\,\sigma^y,\,\sigma^z)$, and $\mathds{1}$ is the $2\times2$ identity matrix. Interestingly, we find that $\textbf{d}^\eta_\theta(\textbf{q})\propto\textbf{q}$ for any direction $\theta$; that is the quasi-particles in the vicinity of such points resemble graphene-like Dirac fermions with a linear dispersion relation. In the main text, we analyze the extreme cases when the strain is applied along the sawtooth ($\theta=0$) and zigzag ($\theta=\pi/2$) directions of the lattice. Now, if we expand the Hamiltonian in the vicinity of the new Dirac points for such cases we find
\begin{eqnarray}
     \text{Sawtooth:}\qquad H_{0}^\eta(\textbf{q})&=&2t_0(1-\beta\epsilon)\mathds{1}-\eta\left(v_x\hbar q_x\sigma^x-v_y\hbar q_y\sigma^z+v_t\hbar q_y\mathds{1}\right)\,,
     \label{saw}\\
      \text{Zigzag:}\qquad H_{\pi/2}^\eta(\textbf{q})&=&2t_0(1-\beta\epsilon)\mathds{1}-\eta(v_y\hbar q_y\sigma^x+v_x\hbar q_x\sigma^z+v_t\hbar q_x\mathds{1})\,,
 \label{zig}
\end{eqnarray}
where the velocities $v_i$ are given in Table \ref{table} as a function of the strain parameters. To first order in strain, we find that $v_x\neq v_y$, meaning that the Dirac cones in the energy spectrum are anisotropic, as well as tilted along the $q_y$ ($q_x$) direction when the strain is applied along the sawtooth (zigzag) direction, as characterized by the tilting velocity $v_t$. From the expressions of the velocities, we also see that when the strain is applied along the sawtooth direction, we have $v_y=v_t$, which means that the Dirac cones are type-III with critical tilting and flat dispersion \cite{type3,weyl}. When the strain is applied along the zigzag direction, we find that the Dirac cones are (tilted) type-I since $v_x>v_t$.

{
\extrarowheight = -0.4ex
\renewcommand{\arraystretch}{2.25}
\begin{table}[]
    \centering
    \begin{tabular}{|l||c|c|c|} 
    \hline 
          & {\large$v_x$} & {\large$v_y$} & {\large$v_t$}   \\ 
         \hline\hline
         $\theta=0$ & $va_0\,Q(\pi/2)\left\{1+\epsilon\left[\beta(3\rho-1)/4+1-\rho\right]\right\}$ & $va_0\,Q(\pi/2)\left\{1+\epsilon\left[\beta(3\rho-1)/4-2\rho\right]\right\}$ & $va_0\,Q(\pi/2)\left\{1+\epsilon\left[\beta(3\rho-1)/4-2\rho\right]\right\}$ \\
        \hline
        $\theta=\pi/2$ & $va_0\,Q(0)\left\{1+\epsilon\left[\beta(1+5\rho)/4-2\rho\right]\right\}$ & $va_0\,Q(0)\left\{1+\epsilon\left[\beta(\rho-3)/4+1-\rho\right]\right\}$ & $va_0\,Q(0)\left\{1+\epsilon\left[\beta(3\rho-1)/4-2\rho\right]\right\}$\\ 
    \hline
    \end{tabular}
    \caption{Velocities in the effective Hamiltonians in Eqs.\ \ref{saw}-\ref{zig} when the strain is applied along the sawtooth ($\theta=0$) and zigzag ($\theta=\pi/2$) directions, respectively. Here we have defined $v=t_0a_0/\hbar$, and $Q(\phi)$ is defined in Eq.\ \ref{Q}.}
    \label{table}
\end{table}
}

\section{Energy gap at the M point}

It can be easily shown that the location of the M point of the strained Brillouin zone is given by
\begin{equation}
    \textbf{M}=\frac{\pi}{2\sqrt{3}\,a_0(\epsilon+1)(\rho\epsilon-1)}\big(\epsilon(1+\rho)\sin2\theta,\,2\epsilon(\rho\sin^2\theta-\cos^2\theta)-2\big)\,.
\end{equation}
If the strain is applied along the sawtooth or zigzag direction, the location of the M point is then given by 

\begin{eqnarray}
    \text{Sawtooth:}\qquad \textbf{M} &=& \left(0,\,\frac{\pi}{\sqrt{3}a_0(1-\rho\epsilon)}\right),\\
    \text{Zigzag:}\qquad \textbf{M} &=& \left(0,\,\frac{\pi}{\sqrt{3}a_0(1+\epsilon)}\right).
\end{eqnarray}
In the presence of different on-site energies and intrinsic SOC, the topological phase transitions occur at the M point as discussed in the manuscript. Such transitions take place when the energy gap $\Delta$ between the upper and middle bands at the M point vanishes, which reads 
\begin{equation}
    \Delta=\left|\varepsilon_{\text{A}}-2\sqrt{\lambda_I^2+t_0^2\left[\beta\sqrt{(\epsilon_{xx}+1)^2+\epsilon_{xy}^2}-\beta-1\right]^2}\right|\,,
    \label{gapM}
\end{equation}
where we have considered $\varepsilon_{\text{A}}\neq0$, $\varepsilon_{\text{B}}=\varepsilon_{\text{C}}=0$ as in the main text. 
The magnitude of the gap is shown in Fig. (4) of the manuscript as a function of $\epsilon$ and $\theta$ for $\varepsilon_{\text{A}}=1.8t_0$ and $\lambda_I=0.1t_0$. The energy gap at the M point for the sawtooth and zigzag strain directions reduces to
\begin{eqnarray}
    \text{Sawtooth:}\qquad \Delta&=&\left|\varepsilon_{\text{A}}-2\sqrt{\lambda_I^2+t_0^2(\beta\epsilon-1)^2}\right|\,,\\
    \text{Zigzag:}\qquad \Delta&=&\left|\varepsilon_{\text{A}}-2\sqrt{\lambda_I^2+t_0^2(\beta\rho\epsilon+1)^2}\right|\,.
\end{eqnarray}

\section{Energy gap at the Dirac points}

The presence of uniform strain shifts also the corners of the Brillouin zone, and the Dirac points near 1/3 filling are relocated. Interestingly, the different on-site energies do not open a gap at the Dirac points here, rather they are only shifted. From the Bloch Hamiltonian \eqref{s1}, we find that the Dirac points to first order in strain are relocated to
\begin{equation}
\textbf{K}^{\pm}_{\epsilon}\approx(\mathds{1}-\hat{\epsilon})\cdot\textbf{K}^{\pm}\pm\textbf{A}-(\textbf{K}^{\pm}\mp \kappa_x\hat{\textbf{x}})
\end{equation}
where $\textbf{K}^{\pm}=(\pm2\pi/3a_0,\,0)$ are the positions of the Dirac points in the unstrained case, $4\cos(a_0\kappa_x)=1+\varepsilon_{\text{A}}/t_0-\sqrt{8+(\varepsilon_{\text{A}}/t_0+1)^2}$ (or $\kappa_x\approx2\pi/3a_0-\varepsilon_{\text{A}}/\left(3\sqrt{3}t_0a_0\right)$ for $\varepsilon_{\text{A}}\ll1$), and 
\begin{equation}
    \textbf{A}=\frac{\beta}{a_0}\left(\frac{\epsilon_{xx}-\epsilon_{yy}}{2\sqrt{3}},\,-\frac{\epsilon_{xy}}{\sqrt{3}}\right)
\end{equation} 
is the strain-induced vector potential \cite{strainKagome}. If we evaluate the Hamiltonian in Eq. \eqref{s1} at  $\textbf{K}^{\pm}_{\epsilon}$ for $\lambda_R=0$, we find that the inclusion of the intrinsic SOC gaps the Dirac cones by $\approx4\sqrt{3}\lambda_I$, {\em independent} of the on-site energies, in contrast to the graphene case \cite{kanez2}. \textcolor{black}{This is due to the spatial inversion symmetry of the lattice, which is broken in the honeycomb lattice when site asymmetries are present.}

\section{First-neighbors intrinsic SOC}

The kagome lattice symmetries allow for an intrinsic spin-orbit interaction between first-neighbors in addition to the SOC to second-neighbors discussed in the main text. In this case, the total intrinsic SOC Hamiltonian is described by
\begin{equation}
    H_{\text{SOC}}(\textbf{k})=i2\sum_{i=1}^3 \left[\lambda\cos(\textbf{k}\cdot{\bm{\delta}}'_i)+\lambda_I\cos(\textbf{k}\cdot\tilde{{\bm{\delta}}}'_i)\right]s^z\otimes A_i
\end{equation}
where $\lambda$ is the intrinsic SOC strength between first-neighbors. 

In the absence of strain and site asymmetries, the effect of this new contribution on the band gaps at the $K$, $K'$ points and the $\Gamma$ point is additive, that is to say, the lower bands are gapped by $2\sqrt{3}(\lambda+2\lambda_I)$ at the Dirac points, while the degeneracy at the $\Gamma$ point between the upper bands is lifted by $4\sqrt{3}(\lambda+\lambda_I)$. On the other hand, $\lambda$ competes with $\lambda_I$ at the M point, such that Eq.\ \eqref{gapM} is modified to
\begin{equation}
\Delta=\left|\varepsilon_{\text{A}}-2\sqrt{(\lambda_I-\lambda)^2+t_0^2\left[\beta\sqrt{(\epsilon_{xx}+1)^2+\epsilon_{xy}^2}-\beta-1\right]^2}\right|\,.
\end{equation}
These rescalings of the effective SOC produce modification in the topological phase diagrams  accordingly.

\bibliography{biblio.bib}